\documentclass[12pt]{article}
\usepackage{epsfig}
\def\be{\begin{equation}}
\def\ee{\end{equation}}
\def\bea{\begin{eqnarray}}
\def\eea{\end{eqnarray}}
\usepackage{graphicx}% Include figure files

\catcode`\@=11
\def\lsim{\mathrel{\mathpalette\@versim<}}
\def\gsim{\mathrel{\mathpalette\@versim>}}
\def\@versim#1#2{\vcenter{\offinterlineskip
\ialign{$\m@th#1\hfil##\hfil$\crcr#2\crcr\sim\crcr } }}
\catcode`\@=12

\catcode`@=12
\begin{document}
\thispagestyle{empty}
\begin{flushright}
UCRHEP-T556\\
November 2015\
\end{flushright}
\vspace{0.6in}
\begin{center}
{\LARGE \bf Neutrino Mixing: A$_4$ Variations\\}
\vspace{1.2in}
{\bf Ernest Ma\\}
\vspace{0.2in}
{\sl Physics \& Astronomy Department and Graduate Division,\\ 
University of California, Riverside, California 92521, USA\\}
\end{center}
\vspace{1.2in}

\begin{abstract}\
In the context of the non-Abelian discrete symmetry $A_4$, the neutrino 
mass matrix has been studied extensively.  A brief update is presented to 
focus on the conceptual shift from tribimaximal mixing ($\theta_{13} = 0$, 
$\theta_{23} = \pi/4$, $\tan^2 \theta_{12} = 1/2$) to \underline{cobimaximal} 
mixing ($\theta_{13} \neq 0$, $\theta_{23} = \pi/4$, $\delta_{CP} = 
\pm \pi/2$) which agrees well with present data.  Three specific realistic 
examples are proposed, two with three and the third with just two 
parameters.
\end{abstract}

\newpage
\baselineskip 24pt

The non-Abelian discrete symmetry $A_4$ is the symmetry of the tetrahedron. 
It has 12 elements and is the smallest group which admits an irreducible 
\underline{3} representation.  It also has three one-dimensional 
representations $\underline{1}, \underline{1}', \underline{1}''$.  
The basic multiplication rule is
\begin{equation}
\underline{3} \times \underline{3} = \underline{1} + \underline{1}' 
+ \underline{1}'' + \underline{3} + \underline{3}. 
\end{equation}
Its application to neutrino mixing began with Ref.~\cite{mr01}, where the 
representation matrices were chosen so that
\begin{eqnarray}
&& a_1 b_1 + a_2 b_2 + a_3 b_3 \sim \underline{1}, \\
&& a_1 b_1 + \omega a_2 b_2 + \omega^2 a_3 b_3 \sim \underline{1}', \\
&& a_1 b_1 + \omega^2 a_2 b_2 + \omega a_3 b_3 \sim \underline{1}'', \\
&& (a_2 b_3 \pm a_3 b_2, a_3 b_1 \pm a_1 b_3, a_1 b_2 \pm a_2 b_1) \sim 
\underline{3}, 
\end{eqnarray}
where $a_i, b_i \sim \underline{3}$ and $\omega = \exp(2 \pi i/3) = -1/2 + 
i \sqrt{3}/2$.  The three lepton families are assumed to transform as 
follows:
\begin{equation}
(\nu_i,l_i)_L \sim \underline{3}, ~~~ l^c_{iL} \sim \underline{1}, 
\underline{1}', \underline{1}'', 
\end{equation}
with three Higgs doublets $(\phi^+_i,\phi^0_i) \sim \underline{3}$. 
Hence the charged-lepton mass matrix is given by
\begin{eqnarray}
{\cal M}_l &=& \pmatrix{f_e v_1^* & f_\mu v_1^* & f_\tau v_1^* \cr 
f_e v_2^* & f_\mu \omega^2 v_2^* & f_\tau \omega v_2^* \cr 
f_e v_3^* & f_\mu \omega v_3^* & f_\tau \omega^2 v_3^*} \nonumber \\ 
&=& \pmatrix{v_1^* & 0 & 0 \cr 0 & v_2^* & 0 \cr 0 & 0 & v_3^*} 
\pmatrix{1 & 1 & 1 \cr 1 & \omega^2 & \omega \cr 1 & \omega & \omega^2}  
\pmatrix{f_e & 0 & 0 \cr 0 & f_\mu & 0 \cr 0 & 0 & f_\tau}.
\end{eqnarray}
For $v_1 = v_2 = v_3$, the $A_4$ symmetry breaks to its residual $Z_3$ and the 
unitary transformation linking ${\cal M}_l$ to ${\cal M}_\nu$ is~\cite{c78,w78}
\begin{equation}
U_\omega = {1 \over \sqrt{3}} \pmatrix{1 & 1 & 1 \cr 1 & \omega & \omega^2 
\cr 1 & \omega^2 & \omega}.
\end{equation}
In the $(e,\mu,\tau)$ basis, the neutrino mass matrix (assumed Majorana) is
\begin{equation}
{\cal M}_\nu^{(e,\mu,\tau)} = U_\omega {\cal M}_A U_\omega^T.
\end{equation}
In general, ${\cal M}_A$ is a $3 \times 3$ symmetric complex matrix, i.e.
\begin{equation}
{\cal M}_A = \pmatrix{a & c & e \cr c & d & b \cr e & b & f}.
\end{equation}
For many years, the neutrino mixing matrix is conjectured to be of 
the tribimaximal form~\cite{hps02}, so that
\begin{equation}
{\cal M}_\nu^{(e,\mu,\tau)} = U_B {\cal M}_B U_B^T,
\end{equation}
where
\begin{equation}
U_B = \pmatrix{\sqrt{2/3} & 1/\sqrt{3} & 0 \cr -1/\sqrt{6} & 1/\sqrt{3} & 
-1/\sqrt{2} \cr -1/\sqrt{6} & 1/\sqrt{3} & 1/\sqrt{2}}.
\end{equation}
If $U_B$ is indeed the correct neutrino mixing matrix, then ${\cal M}_B$ 
would be diagonal.  In general however, it is given by~\cite{m04,mw11}
\begin{equation}
{\cal M}_B = \pmatrix{m_1 & m_6 & m_4 \cr m_6 & m_2 & m_5 \cr m_4 & m_5 & m_3},
\end{equation}
where again $m_{1,2,3,4,5,6}$ are complex.  Nonzero $m_{4,5,6}$ indicate thus 
the deviation from tribimaximal mixing.
The $A_4$ basis is related to the tribimaximal basis through
\begin{equation}
{\cal M}_B = U_A^\dagger {\cal M}_A U_A^*,
\end{equation}
where
\begin{equation}
U_A = U_\omega^\dagger U_B = \pmatrix{0 & 1 & 0 \cr 1/\sqrt{2} & 0 & i/\sqrt{2}    
\cr 1/\sqrt{2} & 0 & -i/\sqrt{2}}.
\end{equation}
Their respective parameters are thus related by
\begin{eqnarray}
&& m_1 = b + (d+f)/2, ~~~ m_2 = a, ~~~ m_3 = b - (d+f)/2, \\ 
&& m_4 = i(f-d)/2, ~~~ m_5 = i(e-c)/\sqrt{2}, ~~~ m_6 = (e+c)/\sqrt{2}.
\end{eqnarray}
To obtain tribimaximal mixing ($\theta_{13} = 0$, $\theta_{23} = \pi/4$, 
$\tan^2 \theta_{12} = 1/2$), $c=e=0$ and $f=d$ are required.  The remaining 
three parameters $(a,b,d)$ are in general complex.  To obtain 
\underline{cobimaximal} mixing ($\theta_{13} \neq 0$, $\theta_{23} = \pi/4$, 
$\delta_{CP} = \pm \pi/2$) which agrees well with present data~\cite{pdg2014} 
with $\delta_{CP} = -\pi/2$~\cite{t2k15}, what is required~\cite{m15} is 
that ${\cal M}_A$ be diagonalized by an orthogonal matrix. To see this, let
\begin{equation}
U_{l \nu} = U_\omega {\cal O},
\end{equation}
where ${\cal O}$ is a real orthogonal matrix, then it is obvious 
that $U_{\mu i} = U^*_{\tau i}$ for $i = 1,2,3$. Comparing this with 
the Particle Data Group (PDG) convention of the neutrino mixing matrix, i.e.
\begin{equation}
U_{l \nu}^{PDG} = \pmatrix{c_{12} c_{13} & s_{12} c_{13} & s_{13} e^{-i \delta} \cr 
-s_{12} c_{23} -c_{12} s_{23} s_{13} e^{i \delta} & c_{12} c_{23} -s_{12} 
s_{23} s_{13} e^{i \delta} & s_{23} c_{13} \cr 
s_{12} s_{23} -c_{12} c_{23} s_{13} e^{i \delta} & -c_{12} s_{23} -s_{12} 
c_{23} s_{13} e^{i \delta} & c_{23} c_{13}}, 
\end{equation}
it is obvious that after rotating the phases of the third column and the 
second and third rows, the two matrices are identical if and only if 
$s_{23} = c_{23}$ and $\cos \delta = 0$, i.e. $\theta_{23} = \pi/4$ and 
$\delta_{CP} = \pm \pi/2$.  This important insight, i.e. Eq.~(18), is a 
rediscovery of what was actually known already many years 
ago~\cite{fmty00,mty01,hs02}.
It is guaranteed if $(a,b,c,d,e,f)$ are all real, so that ${\cal M}_A$ 
is both symmetric and Hermitian.

Another way to arrive at cobimaximal mixing is to use Eqs.~(9) and (10), 
i.e.
\begin{equation}
{\cal M}_\nu^{(e,\mu,\tau)} = U_\omega \pmatrix{a & c & e \cr c & d & b 
\cr e & b & f} U^T_\omega= \pmatrix{A & C & E^* \cr C & D^* & B \cr 
E^* & B & F},
\end{equation}
where
\begin{eqnarray}
A &=& (a + 2b + 2c + d + 2e + f)/3, \\ 
B &=& (a - b - c + d - e + f)/3, \\ 
C &=& (a - b - \omega^2 c + \omega d - \omega e + \omega^2 f)/3, \\ 
D^* &=& (a + 2b + 2 \omega c + \omega^2 d + 2 \omega^2 e + \omega f)/3, \\ 
E^* &=& (a - b - \omega c + \omega^2 d - \omega^2 e + \omega f)/3, \\ 
F &=& (a + 2b + 2 \omega^2 c + \omega d + 2 \omega e + \omega^2 f)/3.
\end{eqnarray}
If again $(a,b,c,d,e,f)$ are real, then $A,B$ are real, whereas $E=C$ 
and $F=D$.  This well-known 
special form was written down already many years ago~\cite{m02,bmv03}, and 
it was pointed out soon afterward~\cite{gl04} that it is protected by a 
generalized $CP$ transformation under $\mu-\tau$ exchange, and it guarantees 
cobimaximal mixing.  With the knowledge that 
$\theta_{13} \neq 0$~\cite{t2k11,dbay12,reno12}, this 
extended symmetry is now the subject of many investigations, which 
began with generalized $S_4$~\cite{mn12}.  In fact, such remnant residual $CP$ 
symmetries are under active study~\cite{cyd15,jp15,hrx15} to reconstruct 
the neutrino mixing matrix with cobimaximal mixing.

Since tribimaximal mixing is not what the data show, ${\cal M}_B$ cannot 
be diagonal.  Many studies are then centered on looking for small 
off-diagonal terms, i.e. $m_{4,5,6}$ which may be complex.  On the other 
hand, data are perfectly consistent with ${\cal M}_A$ as long as it 
is real.  Of course, $\theta_{13}$ and $\theta_{12}$ are not predicted, 
but if extra conditions are imposed, they may be correlated.  For example, 
it has been proposed~\cite{h15} that $c=e=0$, but $f \neq d$, with 
$a,b,d,f$ real for ${\cal M}_A$ in Eq.~(10).  This yields cobimaximal mixing 
together with the prediction that 
\begin{equation}
\tan^2 \theta_{12} = {1 \over 2 - 3 \sin^2 \theta_{13}} > {1 \over 2}.
\end{equation}
Using the 2014 Particle Data Group value~\cite{pdg2014}
\begin{equation}
\sin^2 (2 \theta_{13}) = (9.3 \pm 0.8) \times 10^{-2},
\end{equation}
the value of $\sin^2 (2 \theta_{12})$ from Eq.~(27) is 0.90 with very little 
deviation, as compared with the PDG value
\begin{equation}
\sin^2 (2 \theta_{12}) = 0.846 \pm 0.021,
\end{equation}
which is more than two standard deviations away. 
This is a generic result corresponding to choosing $m_5 = m_6 = 0$ in Eq.~(13).

If $m_4=m_6=0$ is chosen instead, then another generic prediction is
\begin{equation}
\tan^2 \theta_{12} = {1 \over 2} (1 - 3 \sin^2 \theta_{13}).
\end{equation}
Again using Eq.~(28), $\sin^2 (2 \theta_{12}) = 0.866 \pm .002$ is obtained, 
which agrees with Eq.~(29) to within one standard deviation.  Note that both 
generic results hold for 
arbitrary values of $\delta_{CP}$.  

In Ref.~\cite{mw11}, $e+c=0$ is assumed 
so that $m_6=0$.  In addition, $\delta_{CP} = 0$ and $\theta_{23} = \pi/4$ are 
assumed, which can be achieved if both $m_4$ and $m_5$ are nonzero. In the 
case $m_4=m_6=0$, but $m_{1,2,3,5}$ complex, an analysis shows~\cite{im12} 
that large $\delta_{CP}$ correlates with $\theta_{23} \neq \pi/4$ for a fixed 
nonzero $\theta_{13}$.   With the present data, these scenarios are no longer 
favored.  The message now is that cobimaximal mixing should be chosen as 
the preferred starting point of any improved model of neutrino mass and 
mixing.  

Consider a real ${\cal M}_A$ of Eq,~(10) with $d=f$ and $c=-e$, i.e.
\begin{equation}
{\cal M }_A = \pmatrix{a & -e & e \cr -e & d & b \cr e & b & d}.
\end{equation}
In that case,
\begin{equation}
{\cal M}_B = \pmatrix{b+d & 0 & 0 \cr 0 & a & i \sqrt{2} e \cr 0 & 
i \sqrt{2} e & b-d},
\end{equation}
i.e. $m_4=m_6=0$, hence the desirable condition of Eq.~(30) is obtained.
Let ${\cal M}_B$ be diagonalized by
\begin{equation}
U_E = \pmatrix{1 & 0 & 0 \cr 0 & c & is \cr 0 & is & c},
\end{equation}
so that
\begin{equation}
{\cal M}_B = U_E \pmatrix{m'_1 & 0 & 0 \cr 0 & m'_2 & 0 \cr 0 & 0 & m'_3} 
U_E^T,
\end{equation}
where $s = \sin \theta_E$, $c = \cos \theta_E$.  Then
\begin{equation}
{s c \over c^2-s^2} = {e \sqrt{2} \over a+b-d},
\end{equation}
and the three neutrino mass eigenvalues are
\begin{eqnarray}
m'_1 &=& b+d, \\ 
m'_2 &=& {1 \over c^2-s^2} [c^2 a + s^2 (b-d)], \\
m'_3 &=& {1 \over c^2-s^2} [s^2 a + c^2 (b-d)].
\end{eqnarray}
The neutrino mixing matrix is now $U_B U_E$, from which 
\begin{equation}
s = \sqrt{3} \sin \theta_{13}
\end{equation}
is obtained.  As it is, ${\cal M}_A$ has four real parameters $(a,b,d,e)$ 
to fit three observables $(\theta_{13}, \Delta m^2_{21}, \Delta m^2_{32})$, 
hence no prediction is possible other than cobimaximal mixing and Eq.~(30).

In the case of tribimaximal mixing, i.e. $e=0$, the simplest $A_4$ 
model~\cite{af05,m05} has $d=a$.  With this condition, but $e \neq 0$, 
the three neutrino masses are
\begin{equation}
m'_1 = b+a, ~~~ m'_2 = a + {s^2 b \over c^2 -s^2}, ~~~ m'_3 = -a 
+ {c^2 b \over c^2 -s^2}.
\end{equation}
Using Eq.~(28) with the central value $s=0.2673$, they become
\begin{equation}
m'_1 = b+a, ~~~ m'_2 = a + 0.08336 b, ~~~ m'_3 = -a 
+ 1.08336 b.
\end{equation}
Using the central values of~\cite{pdg2014}
\begin{equation}
\Delta m^2_{21} = 7.53 \pm 0.18 \times 10^{-5}~{\rm eV}^2, ~~~ 
\Delta m^2_{32} = 2.44 \pm 0.06 \times 10^{-3}~{\rm eV}^2, 
\end{equation}
the solution is $b/a = -1.714$ and $a = 0.0183$ eV, with $e/a = -0.3642$.  
Using Eq.~(20), the 
effective neutrino mass in neutrinoless double beta decay is predicted to be
\begin{equation}
m_{ee} = |A| = |a + 2b/3| = 2.6 \times 10^{-3}~{\rm eV},
\end{equation}
which is very small, as expected from a normal ordering of neutrino masses, 
and beyond the sensitivity of current and planned experiments.

Another possible three-parameter model is to assume $d=b$, then
\begin{equation}
m'_1 = 2b, ~~~ m'_2 = 1.08336 a, ~~~ m'_3 = 0.08336 a.
\end{equation}
This implies inverted ordering of neutrino masses with $a = 0.0465$ eV, 
$b = 0.0248$ eV, and $e = 0.0099$ eV.  Hence $m_{ee} = |(a + 4b)/3| = 0.0486$ 
eV which is presumably verifiable in the future.

As a third example, consider the following \underline{new remarkable model} 
of just two parameters, with 
$d = -b = 2a$:
\begin{equation}
m'_1 = 0, ~~~ m'_2 = \left({c^2 - 4s^2 \over c^2 - s^2}\right)a = 0.75 a, 
~~~ m'_3 = \left({s^2 - 4c^2 \over c^2 - s^2}\right)a = -4.25 a.
\end{equation}
As a result, $\Delta m^2_{21}/\Delta m^2_{32}$ is predicted to be 0.032, 
in excellent agreement with the experimental value of 0.031. (This is 
a totally new result.) In this case, 
$a = 0.0116$ eV and $m_{ee} = |a/3| = 3.9 \times 10^{-3}$ eV, with 
$e/a = -0.6375$.

In conclusion, it has been pointed out in this paper that $A_4$ is 
intimately related to \underline{cobimaximal} mixing ($\theta_{13} \neq 0$, 
$\theta_{23} = \pi/4$, $\delta_{CP} = \pm \pi/2$) which agrees well with 
present data, and should replace the previously preferred tribimaximal 
mixing pattern.  In particular, a model is proposed with just \underline{two 
real parameters}, with the following predictions:
\begin{eqnarray} 
&& \theta_{23} = \pi/4, ~~~ \delta_{CP} = \pm \pi/2, ~~~ 
\tan^2 \theta_{12} = {1 \over 2} (1 - 3 \sin^2 \theta_{13}), \\ 
&& {\Delta m^2_{21} \over \Delta m^2_{31}} = \left( {1-15\sin^2 \theta_{13} \over 
4- 15 \sin^2 \theta_{13}} \right)^2, ~~~ m_{ee} = 3.9 \times 10^{-3}~{\rm eV}~
({\rm for}~\sin^2 2 \theta_{13} = 0.093),
\end{eqnarray}
which are all well satisfied by present data (except $m_{ee}$ which is yet 
to be measured).

\medskip
This work is supported in part 
by the U.~S.~Department of Energy under Grant No.~DE-SC0008541.

\baselineskip 16pt
\bibliographystyle{unsrt}

\begin{thebibliography}{99}
\bibitem{mr01} E. Ma and G. Rajasekaran, Phys. Rev. {\bf D64}, 113012 (2001).
\bibitem{c78} N. Cabibbo, Phys. Lett. {\bf 72B}, 333 (1978).
\bibitem{w78} L. Wolfenstein, Phys. Rev. {\bf D18}, 958 (1978).
\bibitem{hps02} P. F. Harrison, D. H. Perkins, and W. G. Scott, Phys. Lett. 
{\bf B530}, 167 (2002).
\bibitem{m04} E. Ma, Phys. Rev. {\bf D70}, 031901(R) (2004).
\bibitem{mw11} E. Ma and D. Wegman, Phys. Rev. Lett. {\bf 107}, 061803 (2011).
\bibitem{pdg2014} Particle Data Group, K. A. Olive {\it et al.}, Chin. Phys. 
{\bf C38}, 090001 (2014).
\bibitem{t2k15} K. Abe {\it et al.}, (T2K Collaboration), Phys. Rev. 
{\bf D91}, 072010 (2015).
\bibitem{m15} E. Ma, Phys. Rev. {\bf D92}, 051301(R) (2015).
\bibitem{fmty00} K. Fukuura, T. Miura, E. Takasugi, and M. Yoshimura, 
Phys. Rev. {\bf D61}, 073002 (2000).
\bibitem{mty01} T. Miura, E. Takasugi, and M. Yoshimura, 
Phys. Rev. {\bf D63}, 013001 (2001).
\bibitem{hs02} P. F. Harrison and W. G. Scott, Phys. Lett. {\bf B547}, 219 
(2002).
\bibitem{m02} E. Ma, Phys. Rev. {\bf D66}, 117301 (2002).
\bibitem{bmv03} K. S. Babu, E. Ma, and J. W. F. Valle, Phys. Lett. {\bf B552}, 
207 (2003).
\bibitem{gl04} W. Grimus and L. Lavoura, Phys. Lett. {\bf B579}, 113 (2004).
\bibitem{t2k11} K. Abe {\it et al.}, (T2K Collaboration), Phys. Rev. Lett. 
{\bf 107}, 041801 (2011).
\bibitem{dbay12} F. P. An {\it et al.} (Daya Bay Collaboration), Phys. Rev. 
Lett. {\bf 108}, 171803 (2012).
\bibitem{reno12} S.-B. Kim {\it et al.} (RENO Collaboration), Phys. Rev. 
Lett. {\bf 108}, 191802 (2012).
\bibitem{mn12} R. N. Mohapatra and C. C. Nishi, Phys. Rev. {\bf D86}, 073007 
(2012).
\bibitem{cyd15} P. Chen, C.-Y. Yao, and G.-J. Ding, Phys. Rev. {\bf D92}, 
073002 (2015).
\bibitem{jp15} A. S. Joshipura and K. M. Patel, Phys. Lett. {\bf B749}, 159 
(2015).
\bibitem{hrx15} H.-J. He, W. Rodejohann, and X.-J. Xu, arXiv:1507.03541 
[hep-ph].
\bibitem{h15} X.-G. He, arXiv:1504.01560 [hep-ph].
\bibitem{im12} H. Ishimori and E. Ma, Phys. Rev. {\bf D86}, 045030 (2012).
\bibitem{af05} G. Altarelli and F. Feruglio, Nucl. Phys. {\bf B720}, 64 (2005).
\bibitem{m05} E. Ma, Phys. Rev. {\bf D72}, 037301 (2005).


\end{thebibliography}

\end{document}